%% file: main.tex
\documentclass{article}

\usepackage{colm2024_conference}

\usepackage{footmisc}
\usepackage{booktabs}
\usepackage{graphicx}
\usepackage{enumitem}
\usepackage{wrapfig}
\usepackage{algorithm}
\usepackage{algpseudocode}
\usepackage[T1]{fontenc}

\usepackage{lmodern}
\usepackage{pifont}
\usepackage{amsmath}
\usepackage{amssymb}
\usepackage{mathtools}
\usepackage{amsthm}
\usepackage{bm}

\usepackage{microtype}
\usepackage{colortbl}
\usepackage[utf8]{inputenc}
\definecolor{lightgray}{rgb}{0.9,0.9,0.9}
\usepackage{caption}
\usepackage{subcaption}
\usepackage{xcolor}
\usepackage{setspace}
\usepackage{url}
\usepackage{multirow}

\usepackage{tabularx}
\usepackage{blindtext}
\usepackage{pgfplots}
\pgfplotsset{compat=1.18}
\usepackage{tikz}
\usetikzlibrary{er,positioning,bayesnet}
\usepackage{makecell}
\usepackage{tipa}
\usepackage{siunitx}
\usepackage{nicefrac}
\usepackage{tocloft}
\usepackage{listings}
\usepackage[raster,skins]{tcolorbox} %
\usepackage{xltabular}
\usepackage{adjustbox}
\usepackage{xurl}
\usepackage{marvosym}
\usepackage{algpseudocode}
\usepackage{enumitem}
\usepackage{verbatim}

\definecolor{outlier1}{RGB}{250,252,255}
\definecolor{outlier2}{RGB}{238,244,250}
\definecolor{outlier3}{RGB}{226,236,245}
\definecolor{outlier4}{RGB}{210,225,240}
\definecolor{outlier5}{RGB}{190,210,230}
\definecolor{outlier6}{RGB}{170,195,220}
\definecolor{outlier7}{RGB}{150,180,210}

\definecolor{loss1}{RGB}{250,253,250}
\definecolor{loss2}{RGB}{242,250,240}
\definecolor{loss3}{RGB}{230,245,230}
\definecolor{loss4}{RGB}{215,235,210}
\definecolor{loss5}{RGB}{200,225,200}
\definecolor{loss6}{RGB}{185,215,185}
\definecolor{loss7}{RGB}{170,205,170}

\definecolor{gap_red1}{RGB}{255,245,205}
\definecolor{gap_red2}{RGB}{255,230,190}
\definecolor{gap_red3}{RGB}{255,210,170}
\definecolor{gap_green1}{RGB}{235,250,205}
\definecolor{gap_green2}{RGB}{220,245,190}
\definecolor{gap_green3}{RGB}{200,235,170}

\author{\textbf{iFlytek Spark Audio Team}
}
\title{Spark-Audio 1.0 Technical Report}

\begin{document}

\maketitle

\begin{abstract}
\input{sections/abstract}

\end{abstract}

\section{Introduction}\label{sec:intro}

\input{sections/intro}

\section{Audio Encoder}\label{sec:audio_encoder_pretraining}
\input{sections/encoder}

\section{Audio--LLM Pretraining and Post-Training}\label{sec:audio_llm_training}

\input{sections/pretrain}

\subsection{Post-Training}
\input{sections/SFT}

\input{sections/OPD}

\section{Evaluation}\label{sec:evaluation}

\input{sections/evaluation}

\section{Representation Quality of the Audio Encoder }\label{sec:ablation}

\input{sections/ablation}

\section{Conclusion}\label{sec:conclusion}

\input{sections/conclusion}

\section{Authors}
\input{sections/authors}

\bibliography{custom}
\bibliographystyle{colm2024_conference}



\end{document}

%% file: sections/abstract.tex
This paper introduces Spark-Audio 1.0\footnote{\url{https://iflytekresearch.iflytek.com/experience/spark-audio}}, a speech foundation model trained on domestically produced computing infrastructure. The model accepts text and speech inputs and generates text outputs. It supports multilingual (99 languages) and multi-dialect (202 dialects) speech recognition, speech translation, audio perception and understanding, and both text- and speech-based dialogue, marking a shift from merely ``hearing clearly'' to genuinely ``understanding'' speech. Architecturally, Spark-Audio 1.0 couples a 0.65B-parameter dense audio encoder with a 30B-A3B Mixture-of-Experts (MoE) large language model (LLM). The audio encoder is pretrained with representation-enhanced BEST-RQ and then refined through multi-task supervised training. The audio--LLM is trained through staged pretraining, followed by post-training that combines supervised fine-tuning with on-policy distillation. In total, the model is trained on 13 million hours of audio data together with large-scale text data. We evaluate Spark-Audio 1.0 on mainstream audio and text benchmarks, as well as on in-house test sets drawn from real-world application scenarios. Despite using substantially less audio training data than Qwen3.5-Omni-Flash, Spark-Audio 1.0 achieves comparable performance on most audio and text tasks, outperforms it in multilingual speech recognition, spoken dialogue, and real-world application scenarios, and approaches the performance of larger closed-source models. Nevertheless, there remains room for improvement in instruction following and complex audio reasoning. Finally, the successful training of Spark-Audio 1.0 on domestic hardware, including Huawei Ascend 910B and Hygon BW1000 chips, demonstrates the feasibility of a domestic technical pathway for building speech foundation models.

%% file: sections/intro.tex
Speech is one of the most natural and information-rich channels of human communication, yet conventional automatic speech recognition (ASR) systems have long been optimized for a single objective: converting acoustic signals into verbatim text. Such systems excel at ``hearing clearly'' but discard much of what makes speech meaningful: paralinguistic cues such as emotion, speaker identity, age and gender, and prosodic information that carries intent across languages and dialects, as well as ambient sound events. As intelligent speech technology permeates daily work and life through input methods, translation devices, meeting transcription, customer service, and smart cockpits, the demand has shifted from transcription accuracy alone toward holistic speech understanding. Meeting this demand requires a foundation model that unifies acoustic perception with linguistic reasoning, rather than a collection of narrowly specialized ASR modules.

Recent progress in large language models (LLMs)~\citep{liu2024deepseek, guo2025deepseek, yang2025qwen3,qwenteam2026qwen35,deepseekai2026deepseekv4highlyefficientmilliontoken} has made the speech foundation model paradigm~\citep{chu2023qwen,Chu2024qwen2audio,tang2024salmonn,dinkel2025midashenglm} increasingly attractive: by projecting continuous audio representations into the token space of a powerful LLM, a single model can be trained to perform recognition, translation, audio understanding, and dialogue within one unified framework. Representative multimodal systems such as the Qwen-Omni series~\citep{xu2025qwen3,team2026qwen3} have demonstrated that a single model can maintain strong performance across text, image, audio, and video modalities, achieving state-of-the-art results on a large proportion of audio benchmarks. 
However, the practical deployment of such models faces significant challenges, particularly in the Chinese ecosystem. State-of-the-art speech foundation models are predominantly trained on foreign computing stacks, creating a hard dependency on restricted hardware and software ecosystems. The feasibility of training a competitive speech foundation model end-to-end on purely domestic infrastructure therefore remains an open and strategically important question. At the same time, competitive performance is commonly assumed to require ever-larger training data and model size, which makes training prohibitively expensive and raises doubts about whether models trained on less data can serve real-world applications.

This paper presents Spark-Audio 1.0, a speech foundation model that demonstrates the feasibility of training a competitive model on domestic infrastructure with substantially less data. The model accepts text and audio as input and produces text output, and is designed as a general-purpose base rather than a single-task system. Architecturally, it pairs a 0.65B-parameter dense audio encoder with a 30B-A3B Mixture-of-Experts (MoE) LLM, and is trained on 13 million hours of audio together with large-scale text corpora. Three design decisions distinguish our approach.

First, a semantically grounded, general-purpose audio encoder. The encoder is pretrained in three progressively expanding stages. In the first, self-supervised stage, we revise the BERT-based Speech pre-Training with Random-projection Quantizer (BEST-RQ)~\citep{chiu2022bestrq} framework into a representation-enhanced variant by replacing the input of the quantizer with frozen deep semantic representations from a pretrained multilingual ASR encoder based on the Whisper large-v3 architecture. This injects semantic priors into the prediction targets without introducing an additional trainable tokenizer, and is designed to yield more compact discrete targets and a more discriminative prediction task that strengthens the modeling of fine-grained temporal structure. In this stage, we progressively expand the pretraining corpus from approximately 1.5 million hours to 2.6 million hours and ultimately to 13 million hours. In the second stage, we use large-scale multilingual speech recognition data for supervised training to reinforce the learning of multilingual semantic information. In the third stage, training is extended to multiple perception tasks—namely speech recognition, emotion, gender, age, and sound event recognition—so that the encoder acquires task-agnostic, general-purpose speech representations suitable for the audio-LLM.

Second, a staged training recipe that preserves linguistic competence. Audio--LLM pretraining proceeds in three stages: adapter-only alignment; full-parameter joint optimization of the encoder, adapter, and LLM on large-scale speech, audio, and text data; and long-context extension to 32K tokens with long-form speech and audio tasks. Post-training combines supervised fine-tuning (SFT) with on-policy distillation (OPD), in which a frozen text-only teacher supervises the responses that the model generates from both audio and text inputs. This recipe is designed to avoid the ``cognitive degradation'' on text-centric tasks that commonly afflicts speech-augmented LLMs.

Third, domestic training. Spark-Audio 1.0 is trained on domestic AI accelerators, including Huawei Ascend 910B and Hygon BW1000 chips, from audio encoder pretraining through audio--LLM pretraining to post-training. It establishes a domestic pathway for the entire training pipeline.

We evaluate Spark-Audio 1.0 on both public benchmarks and in-house test sets drawn from real-world application scenarios. The public evaluation covers audio benchmarks (including ASR, speech translation, audio understanding and spoken dialogue task) and text benchmarks (including general knowledge, instruction following, math and code task).  The in-house test sets cover almost the same types of audio tasks as the public benchmarks, together with Chinese ASR under challenging acoustic conditions. Despite using only one-tenth of the training data of the comparably sized Qwen3.5-Omni-Flash, it achieves comparable and, in some cases, superior results on audio tasks, with clear advantages on realistic, application-oriented tasks, and approaches the performance of larger closed-source speech foundation models. Meanwhile, it performs comparably to Qwen3.5-Omni-Flash on text tasks, including general knowledge, mathematics, and coding. Spark-Audio 1.0 still has room for improvement in instruction following and complex audio reasoning.

The contributions of this work are threefold:
\begin{itemize}
    \item \textbf{Broad multilingual and multi-dialect coverage.}
    We present a speech foundation model supporting speech recognition across 99 languages and 202 dialects.

    \item \textbf{Data-efficient training.}
    We show that the representation-enhanced BEST-RQ encoder pretraining and the staged training recipe enable a model trained on approximately one-tenth of the data of a comparably sized counterpart to match or exceed its performance.

    \item \textbf{Domestic training.}
    We demonstrate that a competitive speech foundation model can be trained end to end on domestic infrastructure.

\end{itemize}

The remainder of this paper is organized as follows. Section~\ref{sec:audio_encoder_pretraining} describes the audio encoder pretraining strategy, including the representation-enhanced BEST-RQ framework and the multi-task supervised training stages. Section~\ref{sec:audio_llm_training} describes the pretraining and post-training of Spark-Audio 1.0, covering the staged audio--LLM training recipe together with SFT and OPD. Section~\ref{sec:evaluation} reports evaluation results on public and in-house benchmarks. Section~\ref{sec:ablation} evaluates the representational capability of the Spark-Audio 1.0 audio encoder by comparing it with mainstream audio encoders. Section~\ref{sec:conclusion} concludes the paper with a summary of our findings, limitations, and future work.

%% file: sections/encoder.tex
\subsection{Overview}
\label{subsec:speech_encoder_overview}

\begin{figure*}[t]
    \centering
    \includegraphics[width=\textwidth]{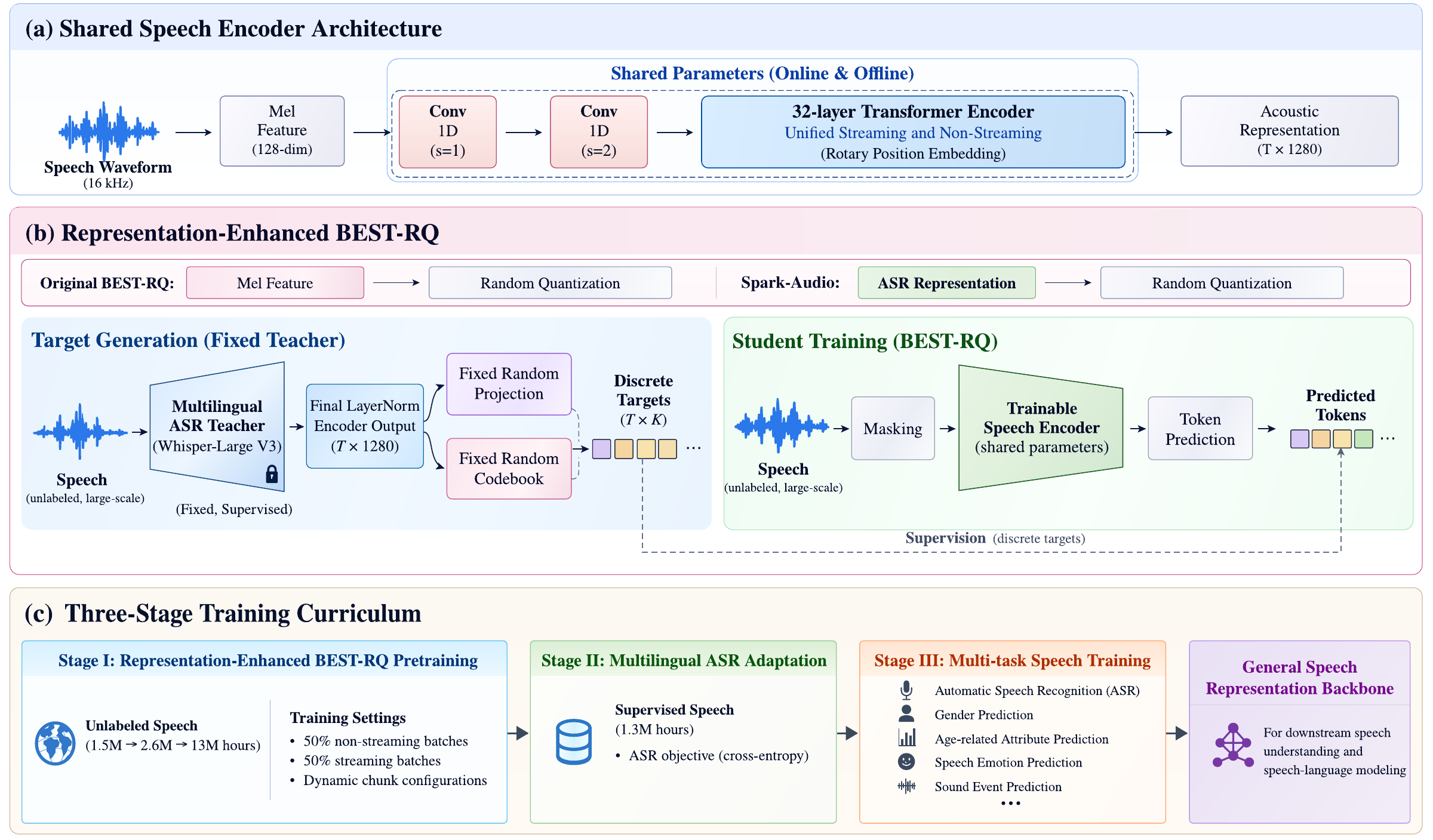}
    \caption{
    Overview of the audio encoder architecture and training curriculum.
    (a) Unified streaming and non-streaming audio encoder based on the Whisper large-v3 architecture.
    (b) Representation-enhanced BEST-RQ pretraining.
    (c) Three-stage training curriculum for the audio encoder.
    }
    \label{fig:speech_encoder_overview}
\end{figure*}

The audio encoder serves as the acoustic front end of Spark-Audio model, encoding input audio into contextual representations of linguistic content and non-linguistic acoustic cues. We adopt the Whisper large-v3 encoder architecture~\citep{radford2023whisper}, with a single set of parameters supporting both streaming and non-streaming inference. As illustrated in Figure~\ref{fig:speech_encoder_overview}, the encoder is trained in three stages. \textbf{Stage~I} performs self-supervised pretraining on up to 13 million hours of unlabeled multilingual audio using a representation-enhanced variant of BEST-RQ~\citep{chiu2022bestrq}. A fixed random-projection quantizer converts contextual features from a frozen multilingual ASR teacher into discrete targets for masked prediction. \textbf{Stage~II} adapts the pretrained encoder to multilingual speech recognition through supervised training. \textbf{Stage~III} introduces multi-task supervision, combining speech recognition with speaker attribute prediction, emotion recognition, and sound event recognition to support broader speech and audio perception.

\subsection{Representation-Enhanced BEST-RQ Pretraining}
\label{subsec:best_rq_pretraining}
To learn general-purpose speech representations from unlabeled data at scale, we build on BEST-RQ~\citep{chiu2022bestrq}, which trains an encoder to predict discrete targets at masked positions. These targets are generated by a random-projection quantizer whose projection matrix and codebook are randomly initialized once and kept fixed throughout pretraining. By avoiding quantizer optimization and iterative clustering, BEST-RQ provides a simple framework for pretraining on large speech corpora.

In the original formulation, the quantizer operates directly on low-level acoustic features. These features reflect linguistic content as well as speaker and recording characteristics, but their representation space is not explicitly shaped by a speech recognition objective. For an encoder intended to support a multilingual speech--language model, we seek to incorporate information learned through multilingual ASR supervision into target construction. By quantizing contextual representations extracted from a pretrained multilingual ASR model, we seek to improve downstream performance and accelerate convergence during encoder pretraining, while retaining the simplicity of fixed random quantization.

To this end, we introduce representation-enhanced BEST-RQ, which applies the random-projection quantizer to contextual representations from a frozen multilingual ASR teacher. The teacher is based on Whisper large-v3 and is trained on approximately 1.5 million hours of multilingual speech using an ASR objective. Given an input utterance $\mathbf{X}$, the teacher encoder produces
\begin{equation}
    \mathbf{Z}
    =
    \mathcal{E}_{\mathrm{T}}(\mathbf{X})
    =
    [\mathbf{z}_{1},\ldots,\mathbf{z}_{T'}],
\end{equation}
where $\mathcal{E}_{\mathrm{T}}$ denotes the teacher encoder, $\mathbf{z}_{t}$ is its representation at time step $t$ after the final encoder layer normalization, and $T'$ is the output sequence length. The teacher and student use the same temporal subsampling configuration, so the teacher representations match the temporal resolution of the student outputs.

Each teacher representation $\mathbf{z}_{t}$ is passed directly to the random-projection quantizer to obtain a discrete target for masked prediction. The teacher parameters, projection matrix, and codebook remain unchanged throughout pretraining, ensuring that the target-generation mapping is independent of student updates. This formulation incorporates knowledge from supervised multilingual ASR into the prediction targets without introducing an additional trainable tokenizer or representation-matching network.

\paragraph{Discrete Target Construction}
Following BEST-RQ~\citep{chiu2022bestrq}, we discretize the
teacher representations using a fixed random-projection quantizer.
Let $\mathbf{A}\in\mathbb{R}^{D_q\times D}$ denote the projection
matrix, where $D$ is the teacher representation dimension and
$D_q$ is the projection dimension. Let
$\mathcal{C}=\{\mathbf{c}_1,\ldots,\mathbf{c}_K\}$ denote a codebook
of $K$ vectors, with $\mathbf{c}_k\in\mathbb{R}^{D_q}$.
Both $\mathbf{A}$ and $\mathcal{C}$ are randomly initialized once
and held fixed throughout pretraining. For each teacher representation $\mathbf{z}_t$, we compute
\begin{equation}
    \begin{aligned}
        \mathbf{u}_t
        &=
        \mathbf{A}\mathbf{z}_t, \\
        q_t
        &=
        \operatorname*{arg\,min}_{k\in\{1,\ldots,K\}}
        \left\lVert
            \operatorname{norm}_2(\mathbf{u}_t)
            -
            \operatorname{norm}_2(\mathbf{c}_k)
        \right\rVert_2, 
    \end{aligned}
    \label{eq:asr_informed_quantization}
\end{equation}
where
$\operatorname{norm}_2(\mathbf{v})
=
\mathbf{v}/\lVert\mathbf{v}\rVert_2$
denotes $\ell_2$ normalization.
Applying this operation framewise yields a sequence of discrete
code indices,
\begin{equation}
    \mathbf{q}
    =
    Q_{\mathrm{RQ}}
    \bigl(\mathcal{E}_{\mathrm{T}}(\mathbf{X})\bigr)
    =
    [q_1,\ldots,q_{T'}],
    \label{eq:teacher_target_sequence}
\end{equation}
where $Q_{\mathrm{RQ}}$ denotes framewise random-projection
quantization. As defined above, $\mathcal{E}_{\mathrm{T}}$
includes the teacher encoder's final layer normalization.

The quantization procedure is retained from BEST-RQ. The change
lies in the representation supplied to the quantizer.
Conventional BEST-RQ constructs targets from low-level acoustic
features, whereas our formulation uses contextual representations
produced by the teacher's 32-layer encoder.
We refer to the resulting labels as
\emph{semantic-informed discrete targets}, reflecting the supervised
multilingual ASR training of the representation source.

\paragraph{Masked Prediction Objective}
Targets are constructed from the unmasked acoustic input
$\mathbf{X}$, while the student receives a masked version of
the same input:
\begin{equation}
    \widetilde{\mathbf{X}}
    =
    \operatorname{Mask}(\mathbf{X}).
\end{equation}
The student encoder produces
\begin{equation}
    \mathbf{H}
    =
    \mathcal{E}_{\theta}
    (\widetilde{\mathbf{X}};\mathbf{M})
    =
    [\mathbf{h}_1,\ldots,\mathbf{h}_{T'}],
\end{equation}
where $\theta$ denotes the encoder parameters and $\mathbf{M}$
specifies the attention visibility pattern for streaming or
non-streaming computation. The same teacher-derived targets
are used in both modes.

A linear prediction head with parameters
$\phi=\{\mathbf{W}_q,\mathbf{b}_q\}$
maps each student representation to a categorical distribution
over the $K$ codebook entries:
\begin{equation}
    p_{\theta,\phi,t}
    \bigl(k\mid\widetilde{\mathbf{X}},\mathbf{M}\bigr)
    =
    \left[
        \operatorname{softmax}
        \left(
            \mathbf{W}_q\mathbf{h}_t+\mathbf{b}_q
        \right)
    \right]_k,
    \label{eq:code_prediction}
\end{equation}
where $[\cdot]_k$ denotes the $k$-th component of the probability
vector.

Let
$\mathcal{I}_{\mathrm{mask}}\subseteq\{1,\ldots,T'\}$
denote the valid positions selected for masked prediction at the
encoder output resolution. These positions are obtained by
mapping the input masking pattern to the encoder time grid,
accounting for temporal subsampling and excluding padding.
The pretraining objective is the average cross-entropy over
these positions:
\begin{equation}
    \mathcal{L}_{\text{BEST-RQ}}
    =
    -
    \frac{1}{|\mathcal{I}_{\mathrm{mask}}|}
    \sum_{t\in\mathcal{I}_{\mathrm{mask}}}
    \log
    p_{\theta,\phi,t}
    \bigl(q_t\mid\widetilde{\mathbf{X}},\mathbf{M}\bigr).
    \label{eq:masked_code_prediction_loss}
\end{equation}

Only the student encoder and prediction head are optimized.
The teacher encoder, projection matrix, and codebook remain
fixed throughout pretraining.
Supervision is provided through discrete code indices:
the objective involves neither regression to teacher hidden
states nor matching of the teacher's ASR output distributions.
Although the teacher is trained on approximately 1.5 million
hours of labeled multilingual speech, transcripts are not
required for the student pretraining corpus.

Table~\ref{tab:bestrq_comparison} summarizes the differences
between the two target-construction schemes.
We retain the same target-generation procedure across
unlabeled pretraining corpora of 1.5 million, 2.6 million,
and 13 million hours.
The data-scaling strategy and the subsequent supervised ASR
and multi-task training stages are described in the following
section.

\begin{table}[t]
    \centering
    \small
    \setlength{\tabcolsep}{4pt}
    \renewcommand{\arraystretch}{1.12}
    \caption{
        Comparison of target construction in BEST-RQ and
        representation-enhanced BEST-RQ.
        Transcripts are not required for student pretraining
        in either formulation. Our method additionally uses
        a teacher trained with multilingual ASR supervision.
    }
    \label{tab:bestrq_comparison}
    \begin{tabularx}{\linewidth}{
        @{}
        >{\raggedright\arraybackslash}p{0.27\linewidth}
        >{\raggedright\arraybackslash}X
        >{\raggedright\arraybackslash}X
        @{}
    }
        \toprule
        \textbf{Component}
        & \textbf{BEST-RQ}
        & \textbf{Representation-enhanced BEST-RQ} \\
        \midrule

        Quantization input
        & Low-level acoustic features
        & Final layer-normalized teacher representations \\

        Teacher encoder
        & None
        & Frozen multilingual Whisper large-v3 \\

        Teacher supervision
        & Not applicable
        & Approx.\ 1.5M hours of multilingual ASR data \\

        Projection matrix
        & Randomly initialized; fixed
        & Randomly initialized; fixed \\

        Codebook
        & Randomly initialized; fixed
        & Randomly initialized; fixed \\

        Prediction targets
        & Discrete code indices
        & Discrete code indices \\

        Transcripts for student pretraining
        & Not required
        & Not required \\

        \bottomrule
    \end{tabularx}
\end{table}

\subsection{Training Curriculum and Data Scaling}
\label{subsec:multi_stage_training}

We train the audio encoder through a three-stage curriculum
that combines large-scale representation pretraining, supervised
multilingual ASR adaptation, and multi-task training with textual
targets. Stage~I learns from unlabeled multilingual audio through
semantic-informed masked prediction. Stage~II directly adapts the
pretrained encoder to multilingual speech recognition using
transcribed speech. Stage~III extends the textual supervision
beyond transcription to descriptions of speaker attributes,
emotion, and sound events.
Table~\ref{tab:speech_encoder_training_stages} summarizes the
training curriculum.

\paragraph{Stage I: Large-scale Representation-enhanced BEST-RQ Pretraining}
We first pretrain the shared streaming/non-streaming encoder
using the representation-enhanced BEST-RQ objective described
in Sec.~\ref{subsec:best_rq_pretraining}.
A frozen multilingual ASR teacher provides contextual
representations of the unmasked input, and a fixed random-projection
quantizer converts its final layer-normalized outputs into discrete
prediction targets. The student encoder receives masked acoustic
inputs and predicts these targets at the designated masked
positions. Transcripts are therefore not required for the student
pretraining corpus, although the target construction incorporates
information learned through the teacher's ASR supervision.

We expand the unlabeled pretraining corpus across three data
scales:
\begin{equation}
    1.5\,\text{M hours}
    \;\longrightarrow\;
    2.6\,\text{M hours}
    \;\longrightarrow\;
    13\,\text{M hours}.
    \label{eq:encoder_pretraining_scales}
\end{equation}
At the largest scale, the corpus contains approximately
13 million hours of audio spanning 99 languages.
The encoder architecture, frozen teacher, random quantizer,
and masked prediction objective remain unchanged across
these data scales.

Joint streaming and non-streaming training is maintained
throughout Stage~I. For each training batch, we select
full-context or streaming computation with equal probability
and update the shared encoder under the selected attention
configuration. This stage provides a common initialization
for subsequent supervised adaptation while exposing the
encoder to both attention configurations during large-scale
pretraining.

\paragraph{Stage II: Multilingual ASR Adaptation}
We adapt the pretrained encoder for multilingual ASR using
approximately 1.3 million hours of transcribed multilingual
speech, drawn from the same data sources as those used
to train the ASR teacher.

The transition from Stage~I to Stage~II changes the form
of supervision:
\begin{equation}
    \underbrace{
        \substack{
            \text{semantic-informed}\\
            \text{masked prediction}
        }
    }_{\text{Stage I}}
    \quad\longrightarrow\quad
    \underbrace{
        \substack{
            \text{direct multilingual}\\
            \text{transcript supervision}
        }
    }_{\text{Stage II}}.
    \label{eq:encoder_supervision_transition}
\end{equation}
In Stage~I, ASR supervision influences the student indirectly
through the teacher-derived discrete targets. In Stage~II,
the pretrained encoder is optimized directly for recovering
the linguistic content of speech.
The streaming and non-streaming configurations continue
to share a single set of encoder parameters.

\paragraph{Stage III: Multi-task Training with Textual Targets}
Although ASR provides direct supervision for linguistic
content, transcripts alone do not explicitly describe
speaker characteristics, emotional expression, or
non-speech sounds. These additional sources of information
are relevant to broader speech and audio understanding,
as reflected in prior work on universal speech
representations and general audio representation learning
\citep{yang2021superb,chen2022wavlm,chen2022unispeechsat,chen2023beats}.

Starting from the ASR-adapted encoder, we extend the training
mixture to five tasks:
\begin{equation}
    \mathcal{T}
    =
    \{
        \mathrm{ASR},
        \mathrm{Gender},
        \mathrm{Age},
        \mathrm{Emotion},
        \mathrm{SoundEvent}
    \}.
    \label{eq:encoder_multitask_set}
\end{equation}
All five tasks use textual target sequences.
For ASR, the target is the speech transcript.
For the additional tasks, the target expresses the relevant
speaker attribute, expressed emotion, or sound event in
natural language (e.g., \emph{The speaker is a young adult.}). 

Together, the three stages establish a progression from
large-scale prediction of teacher-derived discrete targets
to direct multilingual ASR supervision and then to broader
textual supervision of speech and audio.
The curriculum is designed to equip the shared encoder
with representations that support linguistic content,
speaker attributes, paralinguistic cues, and non-speech
acoustic information for downstream speech--language
modeling.

\begin{table*}[t]
    \centering
    \small
    \setlength{\tabcolsep}{4pt}
    \renewcommand{\arraystretch}{1.15}
    \caption{Overview of the three-stage audio encoder
    training curriculum. In Stage~III, speaker attributes,
    emotion, and sound events are expressed as textual targets.}
    \label{tab:speech_encoder_training_stages}
    \begin{tabularx}{\textwidth}{
        @{}
        >{\raggedright\arraybackslash}p{0.08\textwidth}
        >{\raggedright\arraybackslash}p{0.24\textwidth}
        >{\raggedright\arraybackslash}p{0.30\textwidth}
        >{\raggedright\arraybackslash}X
        @{}
    }
        \toprule
        \textbf{Stage}
        & \textbf{Training objective}
        & \textbf{Training data}
        & \textbf{Primary purpose} \\
        \midrule

        Stage I
        & Representation-enhanced BEST-RQ masked prediction
        & 1.5M $\rightarrow$ 2.6M $\rightarrow$ 13M hours
          of unlabeled multilingual audio 
        & Learn multilingual speech representations under
          streaming and non-streaming attention \\

        Stage II
        & Supervised multilingual ASR
        & 1.3M hours of transcribed multilingual speech
        & Adapt the pretrained encoder for multilingual ASR \\

        Stage III
        & Multi-task training with textual targets
        & Audio paired with transcripts or textual
          descriptions of speaker gender, age,
          emotion, and sound events
        & Extend audio-to-text supervision to speaker
          attributes, emotion, and sound events \\

        \bottomrule
    \end{tabularx}
\end{table*}

The encoder described above combines a shared parameterization
for streaming and non-streaming computation with a three-stage
training curriculum. Representation-enhanced BEST-RQ pretraining
on large-scale unlabeled multilingual audio is followed by
supervised multilingual ASR adaptation and multi-task training
with textual targets. This curriculum is designed to learn
contextual representations that capture linguistic content
alongside speaker attributes, emotional cues, and non-speech
acoustic information. The resulting encoder serves as the
acoustic front end of our audio-LLM. In the following section,
we describe its integration with the language model and the
pretraining and post-training procedures for the complete
audio--language system.

%% file: sections/pretrain.tex
\subsection{Pretraining}

We combine the pretrained audio encoder with a 30B-A3B MoE language backbone pretrained on large-scale text corpora. The audio encoder produces representations at \SI{50}{\hertz}. An adapter module, consisting of an MLP followed by CNN-based temporal downsampling layers, transforms these representations into a \SI{12.5}{\hertz} sequence for input to the language model.

Our pretraining employs a three-stage curriculum: adapter alignment, full-parameter joint training, and long-context extension. The curriculum is designed to support multilingual and dialectal speech recognition, speech translation, audio understanding, and long-context modeling, while maintaining the language backbone's text capabilities.

\subsubsection{Pretraining Data}
The pretraining corpus includes ASR for Chinese, Chinese dialects, English, and other languages; speech translation; audio understanding; text translation; and general text. The corpus covers 99 languages and 202 dialects across diverse application domains and acoustic conditions, including noisy and low-volume speech.

We formulate all tasks as conditional text generation. Depending on the task, the conditioning inputs include audio, a task prompt, text, and optional contextual information. The model autoregressively generates the corresponding target text. For ASR, prompts specify the transcription language; for speech translation, they specify the source and target languages; and for Chinese-dialect ASR, they specify both the language and the required output format.

We organize the data by task, language, source, annotation quality, and duration, and adjust the training mixture across stages. Stage~I uses a relatively small amount of supervised data for adapter alignment. Stage~II primarily uses large-scale weakly supervised data, supplemented with high-quality supervised data. Stage~III primarily uses supervised data, with greater emphasis on long-form audio, long text, and context-dependent tasks.

\subsubsection{Multi-Stage Training}
Our training curriculum comprises three successive stages that progressively expand the scope of optimization, data coverage, and context length. It begins with adapter alignment, proceeds to large-scale full-parameter training for general capability acquisition, and concludes with long-context extension using an increased proportion of high-quality and long-form data.

Table~\ref{tab:three-stage-recipe} summarizes the training recipe at each stage. All stages use BF16 arithmetic~\citep{kalamkar2019bfloat16} and the AdamW optimizer with $\beta_1=0.9$, $\beta_2=0.95$, $\epsilon=10^{-8}$, and a weight decay of 0.01. We clip the gradient norm at 1 and use a label-smoothing coefficient of 0.05.

\begin{table}[t]
  \centering
  \caption{Training configurations for the three pretraining stages. The global batch size is measured as the total number of input tokens per optimizer update.}
  \label{tab:three-stage-recipe}
  \small
  \setlength{\tabcolsep}{4pt}
  \begin{tabularx}{\textwidth}{@{}lXXX@{}}
    \toprule
     & \textbf{Stage I} & \textbf{Stage II} & \textbf{Stage III} \\
    \midrule
    Trainable components
      & Adapter
      & All
      & All \\
    Learning rate
      & Peak $2\times10^{-4}$
      & $5\times10^{-5} \rightarrow 2.5\times10^{-5}$
      & $2.5\times10^{-5} \rightarrow 5\times10^{-6}$ \\
    Schedule
      & Triangular
      & Cosine decay
      & Cosine decay \\
    Global batch size
      & 67M
      & 66M
      & 152M \\
    Maximum sequence length
      & 4K
      & 4K
      & 32K \\
    Maximum audio chunk duration$^{*}$
      & \SI{30}{\second}
      & \SI{30}{\second}
      & \SI{45}{\second} \\
    Parallelism
      & EP8
      & EP8, ZeRO-2
      & EP8, CP4, ZeRO-2 \\
    \bottomrule
  \end{tabularx}

  \par\smallskip
  \begin{minipage}{\textwidth}
    \footnotesize
    \textit{Note:} The adapter includes both the MLP and the CNN-based
    temporal downsampling layers. EP and CP denote expert parallelism
    and context parallelism~\citep{nvidia2026contextparallelism},
    respectively. ZeRO-2 denotes Stage 2 of the Zero Redundancy
    Optimizer~\citep{rajbhandari2020zero}.
    $^{*}$Recordings longer than this limit are split into chunks
    and encoded independently. The resulting audio embeddings are
    concatenated in temporal order within each sample.
  \end{minipage}
\end{table}

\paragraph{Stage I: Encoder Alignment}
We freeze both the audio encoder and the MoE language backbone and train only the adapter, including its MLP and CNN-based downsampling layers. This stage primarily uses ASR and audio understanding data to align audio representations with the language model's input space. We exclude reasoning-oriented data and focus on establishing alignment between the audio encoder and the LLM before joint training.

\paragraph{Stage II: Large-Scale Joint Training}
We unfreeze all components and jointly optimize the audio encoder, adapter, and language backbone. This stage primarily uses weakly supervised data spanning diverse languages and acoustic conditions, supplemented with high-quality supervised samples. The audio tasks include multilingual and dialectal speech recognition, speech translation, and audio understanding. The data cover diverse domains, recording conditions, and annotation quality levels, including noisy and low-volume speech.

The text-only corpus comprises general-domain, translation, science and mathematics content, multilingual, and code data. We retain text-only data during joint training to support continued learning of text capabilities alongside speech and audio tasks.

\paragraph{Stage III: Long-Context Extension}
We continue joint optimization of all components while extending the maximum sequence length from 4K to 32K tokens. We increase the proportion of high-quality speech data and introduce long-form tasks, including long-form speech recognition, long-form audio understanding, and speech recognition conditioned on textual context. The text-only mixture places greater emphasis on long documents, science and mathematics content, and code.

We also increase the maximum audio chunk duration from \SI{30}{\second} to \SI{45}{\second}. Longer recordings are split into chunks, encoded independently, and concatenated in temporal order at the embedding level. This stage is designed to support tasks that require integrating information across longer audio and text sequences.

\subsubsection{Batch Construction and Long-Context Computation}

To preserve the desired mixture across data sources, we first stratify the training data by source and then construct local batch sequences whose source proportions approximate the prescribed mixture. We deliberately avoid sorting samples by audio length to improve distributional balance and training stability.

Because microbatches can contain different numbers of target tokens, we normalize the training loss by the number of valid target tokens. Let $m_{b,t}\in\{0,1\}$ indicate whether position $t$ in sample $b$ contributes to the loss, and let $\ell_{b,t}$ denote the corresponding label-smoothed cross-entropy loss. The objective for each optimizer update is
\begin{equation}
  \mathcal{L}
  =
  \frac{\sum_{b,t} m_{b,t}\ell_{b,t}}
       {\sum_{b,t} m_{b,t}},
  \label{eq:token-normalized-loss}
\end{equation}
where the sums cover all valid target tokens contributing to one optimizer update, across gradient-accumulation steps and data-parallel workers.

For long-context training, Stage~III combines eight-way expert parallelism (EP) with four-way context parallelism (CP). Audio chunks are first distributed across CP ranks along the batch dimension, and each rank encodes its assigned chunks. The resulting audio embeddings are combined with text-token embeddings to form the complete multimodal sequence, which is then partitioned along the sequence dimension for context-parallel computation~\citep{nvidia2026contextparallelism}.

%% file: sections/SFT.tex
\subsubsection{Supervised Fine-Tuning}

Following pretraining, we perform SFT using high-quality multi-task data organized in ChatML format. This stage aims to consolidate the model's fundamental speech and language capabilities while enhancing its ability to understand complex instructions, perform cognitive reasoning, and engage in natural interactions. To this end, we construct a diverse SFT dataset covering fundamental audio tasks, cognitive tasks, and interaction tasks.

For fundamental audio tasks, we continue training the model on ASR, translation, and audio understanding. We carefully select a high-quality subset of the data employed in the third stage of pretraining to further strengthen the model's existing capabilities. Beyond these fundamental tasks, we introduce additional supervised data to enhance the model's cognitive and interactive abilities. For cognitive tasks, we adopt two complementary supervision formats: text-to-text (T2T) and speech-to-text (S2T). The T2T data encompass reasoning, mathematics, and knowledge-based tasks, while the S2T data are constructed by converting the corresponding text instructions into spoken inputs using text-to-speech (TTS) synthesis. By providing supervision across both modalities, this design aims to narrow the gap in instruction-following and reasoning capabilities between speech and text inputs. We also incorporate interaction data to improve the naturalness and expressiveness of the model's responses.

With this diverse multi-task supervision, we train the model for one epoch, employing a cosine learning rate decay schedule that gradually reduces the learning rate from \(1 \times 10^{-5}\) to \(1 \times 10^{-6}\).

%% file: sections/OPD.tex
\subsubsection{On-Policy Distillation}
\label{sec:opd}

\paragraph{Dual-Trajectory Formulation}
Following pretraining and supervised fine-tuning, we apply dual-trajectory on-policy distillation (OPD) to transfer reasoning and instruction-following behavior from a frozen text-only teacher to the audio--language student. The two trajectories use the same student model under audio and text conditioning, respectively. The audio-conditioned trajectory transfers teacher supervision to spoken inputs, whereas the text-conditioned trajectory provides continued supervision for the student's text-based capabilities. This joint design aims to improve audio-conditioned response generation while limiting degradation in text-conditioned performance.

\paragraph{Student Rollouts and Teacher Conditioning}
Let $\mathcal{B}=\{(S_b,T_b)\}_{b=1}^{B}$ denote a minibatch of paired inputs, where $S_b$ is a spoken input and $T_b$ is its corresponding transcript. We write $x_b^{a}=S_b$ and $x_b^{t}=T_b$ for the two input modalities; shared task instructions and dialogue context are implicit in the conditioning inputs. For each pair, the student generates a separate textual response under each modality:
\begin{equation}
    \mathbf{y}_b^{m}
    \sim \rho_{\theta}(\cdot \mid x_b^{m}),
    \qquad m\in\{a,t\},
    \label{eq:opd_rollouts}
\end{equation}
where $\rho_{\theta}$ denotes the decoding distribution induced by the current student and the sampling settings specified below. Both trajectories produce text; their superscripts identify the input modality.

The teacher evaluates each trajectory using the transcript and the corresponding student-generated response prefix. At response position $i$, the student and teacher logits for token $v$ are
\begin{equation}
\begin{aligned}
    z_{\theta,b,i}^{m}(v)
    &= z_{\theta}(v \mid x_b^{m},\mathbf{y}_{b,<i}^{m}),\\
    z_{\phi,b,i}^{m}(v)
    &= z_{\phi}(v \mid T_b,\mathbf{y}_{b,<i}^{m}),
\end{aligned}
\label{eq:opd_logits}
\end{equation}
where $\theta$ and $\phi$ denote the student and frozen teacher parameters, respectively. The teacher therefore conditions on text in both trajectories, but its next-token distributions are evaluated separately because the two student-generated prefixes may differ. This exposes the student to teacher supervision at prefixes visited by its own decoding policy.

\paragraph{Distillation over Teacher-Selected Support}
Following \citet{fu2026revisitingonpolicydistillationempirical}, we compute a local reverse-KL objective on a teacher-selected token set. Let $\mathcal{V}$ denote the shared text vocabulary with aligned token identities between the two models. For each trajectory and response position, we select
\begin{equation}
    \mathcal{K}_{b,i}^{m}
    =
    \operatorname{TopK}_{K_{\mathrm{KD}}}
    \left(
        \{z_{\phi,b,i}^{m}(v)\}_{v\in\mathcal{V}}
    \right),
    \label{eq:opd_support}
\end{equation}
where $\operatorname{TopK}_{K_{\mathrm{KD}}}$ returns the identities of the $K_{\mathrm{KD}}$ tokens with the highest teacher logits.

For $v\in\mathcal{K}_{b,i}^{m}$, we independently normalize the student and teacher logits on this same support using the distillation temperature $\tau_{\mathrm{KD}}$:
\begin{equation}
\begin{aligned}
    \widetilde{p}_{\theta,b,i}^{m}(v)
    &=
    \frac{
        \exp(z_{\theta,b,i}^{m}(v)/\tau_{\mathrm{KD}})
    }{
        \sum_{u\in\mathcal{K}_{b,i}^{m}}
        \exp(z_{\theta,b,i}^{m}(u)/\tau_{\mathrm{KD}})
    },\\
    \widetilde{q}_{\phi,b,i}^{m}(v)
    &=
    \frac{
        \exp(z_{\phi,b,i}^{m}(v)/\tau_{\mathrm{KD}})
    }{
        \sum_{u\in\mathcal{K}_{b,i}^{m}}
        \exp(z_{\phi,b,i}^{m}(u)/\tau_{\mathrm{KD}})
    }.
\end{aligned}
\label{eq:opd_local_distributions}
\end{equation}
The per-position distillation loss is
\begin{equation}
\begin{aligned}
    d_{b,i}^{m}
    &=
    D_{\mathrm{KL}}(
        \widetilde{p}_{\theta,b,i}^{m}
        \,\Vert\,
        \widetilde{q}_{\phi,b,i}^{m})  \\
    &=
    \sum_{v\in\mathcal{K}_{b,i}^{m}}
    \widetilde{p}_{\theta,b,i}^{m}(v)
    \log
    \frac{
        \widetilde{p}_{\theta,b,i}^{m}(v)
    }{
        \widetilde{q}_{\phi,b,i}^{m}(v)
    }.
\end{aligned}
\label{eq:opd_local_kl}
\end{equation}
This objective matches the conditional distributions within the selected support. It limits the distribution comparison to $K_{\mathrm{KD}}$ candidates per position and does not equal the full-vocabulary reverse KL.

\paragraph{Joint Optimization}
Let $n_b^{m}$ denote the length of response $\mathbf{y}_b^{m}$, and let $M_{b,i}^{m}\in\{0,1\}$ indicate whether a response position contributes to the loss. Padding and designated control-token positions are masked; input-prompt positions are not included in the response loss. We normalize each trajectory loss by its own number of valid response tokens:
\begin{equation}
    \mathcal{L}_{m}
    =
    \frac{
        \sum_{b=1}^{B}\sum_{i=1}^{n_b^{m}}
        M_{b,i}^{m}d_{b,i}^{m}
    }{
        \sum_{b=1}^{B}\sum_{i=1}^{n_b^{m}}
        M_{b,i}^{m}
    },
    \qquad m\in\{a,t\}.
    \label{eq:opd_branch_loss}
\end{equation}
The joint objective is
\begin{equation}
    \mathcal{L}_{\mathrm{OPD}}
    =
    \mathcal{L}_{a}
    +
    \lambda\mathcal{L}_{t},
    \label{eq:opd_joint_loss}
\end{equation}
with $\lambda=1$, giving equal coefficients to the two separately normalized losses. The teacher remains frozen, and the sampled trajectories are treated as fixed during each gradient update. Only the student's trainable parameters are updated through the distillation objective.

\paragraph{Training Configuration}
We use $K_{\mathrm{KD}}=64$ and $\tau_{\mathrm{KD}}=0.5$ for distillation. Student rollouts are generated with temperature $\tau_{\mathrm{gen}}=0.5$, a nucleus-sampling threshold of $p_{\mathrm{gen}}=0.9$, a top-$k$ sampling cutoff of $k_{\mathrm{gen}}=50$, and a repetition penalty of $1.1$. These sampling settings define $\rho_{\theta}$ in Eq.~\eqref{eq:opd_rollouts} and are separate from the support size and temperature used in the distillation loss.

Each optimization step combines the audio- and text-conditioned losses. We use a learning rate of $5\times10^{-6}$ with cosine decay, a warmup ratio of $0.5\%$, a gradient-accumulation factor of $1$, and gradient clipping. The maximum sequence length is $8{,}192$ tokens. Training runs for three epochs, and the reported results use the checkpoint obtained at the end of the second epoch.

\paragraph{Training Scope and Transfer}
This stage uses tens of thousands of paired spoken and textual inputs for reasoning and instruction following. No additional examples with ASR or general audio-perception targets are introduced during this stage. Nevertheless, we observe modest improvements on a subset of recognition and audio-perception evaluations. These gains may reflect improved use of linguistic information under audio conditioning, together with better task interpretation and response generation.

%% file: sections/evaluation.tex

We evaluate Spark-Audio 1.0 across speech recognition, speech-to-text translation, audio understanding, spoken interaction, and text-only tasks. Public benchmarks provide comparisons across a broad range of capabilities, while in-house evaluations assess performance in the product and developer scenarios targeted by the system. We additionally evaluate Chinese ASR under challenging acoustic conditions. Tables~\ref{tab:spark-audio-public-audio}--\ref{tab:spark-audio-broader-capability} summarize the results.

\subsection{Evaluation Setup}
\label{subsec:evaluation_setup}

\paragraph{Public Audio Benchmarks}
We evaluate multilingual ASR on FLEURS~\citep{conneau2022fleurs}. Additional Chinese ASR benchmarks include AISHELL6-Whisper~\citep{li2025aishell6}, which contains whispered and parallel normally voiced Mandarin speech; the internet and meeting subsets of WenetSpeech~\citep{zhang2022wenetspeech}; and SpeechIO~\citep{speechcolab2024speechio}. KeSpeech~\citep{tang2021kespeech} provides a complementary evaluation of Mandarin and regional speech varieties. English ASR is evaluated on FLEURS-en, LibriSpeech~\citep{panayotov2015librispeech}, speechocean762~\citep{zhang2021speechocean762}, and the TED~\citep{hernandez2018tedlium3} evaluation set, covering read speech, non-native speech, and public talks. We also report average FLEURS error rates over three language groups (14, 20, and 64 languages; see the footnotes of Table~\ref{tab:spark-audio-public-audio}).

Beyond transcription, we evaluate audio understanding on MMAR~\citep{ma2025mmar}, AIR-Bench~\citep{yang2024airbench}, and MMSU~\citep{wang2025mmsu}, and spoken interaction on VoiceBench~\citep{chen2024voicebench}. Speech-to-text translation (S2TT) is evaluated in both Chinese-to-English and English-to-Chinese directions using FLEURS.

\paragraph{In-house Audio Evaluations}
Our in-house evaluations cover general Chinese ASR, a 202-dialect recognition corpus, multilingual ASR, spoken interaction, and Chinese--English S2TT. The evaluation data are drawn from real-world application scenarios. Multilingual ASR results are reported for a 14-language group and an 18-language group. Spoken interaction is evaluated on general interaction and instruction following. A separate Chinese ASR evaluation focuses on high-noise and low-volume speech to characterize performance under challenging acoustic conditions.


\paragraph{Text-only Benchmarks}
We assess text-based knowledge with MMLU-Pro~\citep{wang2024mmlupro}, CMMLU~\citep{li2023cmmlu}, and C-Eval~\citep{huang2023ceval}; verifiable instruction following with IFEval~\citep{zhou2023ifeval} and IFBench~\citep{pyatkin2025ifbench}; mathematical problem solving with MATH-500~\citep{hendrycks2021math} and GSM8K~\citep{cobbe2021gsm8k}; and code generation with HumanEval~\citep{chen2021humaneval}. These evaluations characterize the text capabilities of the final audio--language model alongside its speech and audio performance.

\paragraph{Metrics and Comparison Systems}
ASR performance is measured using word error rate (WER) or character error rate (CER), expressed as percentages, and S2TT performance is measured using BLEU. Public audio understanding and spoken interaction results are reported using the metrics indicated in Table~\ref{tab:spark-audio-public-audio}. For in-house spoken interaction, a DeepSeek judge assigns each response a score from 0 to 5. We report the \emph{full-score rate} (FSR), defined as the percentage of responses receiving the maximum score:
\begin{equation}
    \mathrm{FSR}
    =
    \frac{100}{N}
    \sum_{i=1}^{N}
    \mathbf{1}\{s_i = 5\},
    \label{eq:dialogue_full_score_rate}
\end{equation}
where $N$ is the number of evaluated responses and $s_i$ is the judge score assigned to response $i$.

The audio evaluations compare Spark-Audio 1.0 with Qwen3.5-Omni-Flash, Qwen3.5-Omni-Plus, and Gemini 3.1 Pro where results are available. The text-only evaluation compares Spark-Audio 1.0 with Qwen3.5-Omni-Flash. All comparative statements below refer to the systems and evaluation subsets reported in the corresponding tables.

\begin{table*}[!t]
\centering
\caption{Performance on public audio benchmarks. ASR is evaluated using WER or CER (\%), and speech-to-text translation (S2TT) using BLEU. Audio understanding and spoken interaction adopt official evaluation methods and metrics. Lower WER/CER and higher values of the other metrics are better. Bold indicates the best result among the systems reported for each evaluation; -- denotes an unreported result.}
\label{tab:spark-audio-public-audio}
\footnotesize
\setlength{\tabcolsep}{2pt}
\renewcommand{\arraystretch}{1.08}
\begin{tabularx}{\linewidth}{@{}
>{\raggedright\arraybackslash}p{0.17\linewidth}
>{\raggedright\arraybackslash}p{0.23\linewidth}
>{\centering\arraybackslash}p{0.10\linewidth}
*{4}{>{\centering\arraybackslash}X}@{}}
\toprule
\textbf{Task} & \textbf{Benchmark / set} & \textbf{Metric} & \shortstack[c]{\textbf{Spark-Audio}\\\textbf{1.0}} & \shortstack[c]{\textbf{Qwen3.5-}\\ \textbf{Omni-Flash}} & \shortstack[c]{\textbf{Qwen3.5-}\\ \textbf{Omni-Plus}} & \shortstack[c]{\textbf{Gemini 3.1}\\\textbf{Pro}} \\
\midrule
\multirow{4}{=}{\raggedright Chinese ASR} & FLEURS-zh & CER $\downarrow$ & \textbf{1.88} & 2.90 & 2.90 & 3.60 \\
 & AISHELL6-Whisper & CER $\downarrow$ & 2.63 & 3.04 & \textbf{2.36} & 14.62 \\
 & \shortstack[l]{WenetSpeech (net | meeting)} & CER $\downarrow$ & \shortstack[r]{4.97 | 6.38} & \shortstack[r]{4.41 | \textbf{5.51}} & \shortstack[r]{\textbf{4.30} | 5.84} & \shortstack[r]{11.53 | 14.21} \\
 & SpeechIO & CER $\downarrow$ & 3.15 & 4.77 & \textbf{2.87} & -- \\
\midrule
Dialect ASR & KeSpeech & CER $\downarrow$ & 4.71 & 4.47 & \textbf{3.46} & 23.67 \\
\midrule
\multirow{4}{=}{\raggedright English ASR} & FLEURS-en & WER $\downarrow$ & \textbf{2.25} & 3.70 & 3.20 & 2.70 \\
 & \shortstack[l]{LibriSpeech (clean | other)} & WER $\downarrow$ & \shortstack[r]{\textbf{1.08} | 2.44} & \shortstack[r]{1.30 | 2.43} & \shortstack[r]{1.11 | \textbf{2.23}} & \shortstack[r]{3.36 | 4.41} \\
 & speechocean762 & WER $\downarrow$ & \textbf{10.14} & 10.80 & 10.43 & 16.58 \\
 & TED & WER $\downarrow$ & \textbf{7.13} & 11.29 & 11.10 & 8.33 \\
\midrule
\multirow{3}{=}{\raggedright Multilingual ASR} & FLEURS (core)\textsuperscript{a} & \mbox{WER/CER $\downarrow$} & \textbf{2.81} & 3.96 & 3.47 & 3.36 \\
 & FLEURS (extended)\textsuperscript{b} & \mbox{WER/CER $\downarrow$} & \textbf{6.30} & 12.08 & 6.65 & 8.38 \\
 & FLEURS (additional)\textsuperscript{c} & \mbox{WER/CER $\downarrow$} & 31.89 & 39.00 & 28.66 & \textbf{20.09} \\

\midrule
\multirow{3}{=}{\raggedright Audio understanding} & AIR-Bench & Score $\uparrow$ & 71.85 & 72.63 & \textbf{79.23} & -- \\
& MMSU & Acc. $\uparrow$ & 73.96 & 72.20 & \textbf{82.80} & 81.30 \\
& MMAR & Acc. $\uparrow$ & 57.90 & 74.00 & 80.00 & \textbf{83.70} \\

\midrule
Spoken interaction & VoiceBench & Score $\uparrow$ & 90.35 & 87.80 & \textbf{93.10} & 88.90 \\
\midrule
\multirow{2}{=}{\raggedright S2TT} & FLEURS zh$\rightarrow$en & BLEU $\uparrow$ & 28.40 & 31.20 & \textbf{32.20} & 30.10 \\
 & FLEURS en$\rightarrow$zh & BLEU $\uparrow$ & 44.54 & 46.60 & \textbf{47.80} & 47.40 \\
\bottomrule
\end{tabularx}
\vspace{2pt}
\begin{minipage}{\linewidth}
\scriptsize
\setlength{\parindent}{0pt}
\textsuperscript{a}Core group (14 languages): Mandarin Chinese, English, Japanese, Spanish, Arabic, German, French, Italian, Portuguese, Korean, Russian, Hindi, Thai, and Vietnamese.\par
\textsuperscript{b}Extended group (20 languages): Bulgarian, Bengali, Czech, Greek, Persian, Filipino, Indonesian, Kazakh, Mongolian, Malay, Dutch, Polish, Romanian, Swedish, Swahili, Tamil, Turkish, Ukrainian, Urdu, and Uzbek. Hausa is excluded from this reported FLEURS aggregate because anomalous results were observed for some comparison systems.\par
\textsuperscript{c}Additional group (64 languages): Cantonese, Afrikaans, Amharic, Assamese, Asturian, Azerbaijani, Belarusian, Bosnian, Catalan, Cebuano, Welsh, Danish, Estonian, Fula, Finnish, Irish, Galician, Gujarati, Hebrew, Croatian, Hungarian, Armenian, Igbo, Icelandic, Javanese, Kamba, Kabuverdianu, Khmer, Kannada, Kyrgyz, Luxembourgish, Luganda, Lingala, Lao, Lithuanian, Luo, Latvian, Maori, Macedonian, Malayalam, Marathi, Maltese, Burmese, Norwegian, Nepali, Northern Sotho, Nyanja, Occitan, Oromo, Punjabi, Pashto, Sindhi, Slovak, Slovenian, Shona, Somali, Serbian, Telugu, Tajik, Umbundu, Wolof, Xhosa, Yoruba, and Zulu. Sorani Kurdish is excluded from this aggregate because anomalous results were observed for some comparison systems.
\end{minipage}
\end{table*}

\subsection{Speech Recognition}
\label{subsec:asr_evaluation}

\paragraph{Public Benchmarks}
As shown in Table~\ref{tab:spark-audio-public-audio}, Spark-Audio 1.0 achieves the lowest reported error rates on FLEURS Chinese and English, with a CER of 1.88\% and a WER of 2.25\%, respectively. Relative to the best comparison result on each subset, these correspond to error-rate reductions of 35.2\% and 16.7\%. It also achieves the lowest reported WER on the LibriSpeech clean subset (1.08\%), speechocean762 (10.14\%), and TED (7.13\%), demonstrating competitive recognition across several English speech conditions. Performance is more mixed on the remaining benchmarks: Qwen3.5-Omni-Plus performs better on AISHELL6-Whisper and SpeechIO, while both Qwen variants obtain lower error rates on KeSpeech, the LibriSpeech other subset, and the WenetSpeech internet and meeting subsets.

The multilingual results also vary across language groups. Spark-Audio 1.0 obtains the lowest reported average error rates for the 14-language and 20-language FLEURS groups, at 2.81\% and 6.30\%, respectively. On the 64-language group, its average error rate of 31.89\% is lower than that of Qwen3.5-Omni-Flash (39.00\%) but higher than those of Qwen3.5-Omni-Plus (28.66\%) and Gemini 3.1 Pro (20.09\%). The results therefore show an advantage on the first two evaluated language groups, while identifying broader multilingual recognition as an important area for improvement.

\paragraph{In-house Recognition}
Spark-Audio 1.0 achieves the lowest reported error rate on each ASR evaluation in Table~\ref{tab:spark-audio-inhouse-audio}. On general Chinese recognition, it obtains a CER of 7.26\%, compared with 7.45\% for the best comparison system. The margin is larger on the 202-dialect corpus, where its CER of 14.29\% improves on the best comparison result of 19.79\% by 5.50 percentage points, corresponding to a 27.8\% relative reduction. For the 14-language and 18-language groups, its average error rates are 7.75\% and 9.87\%, compared with 10.86\% and 12.06\% for the respective best comparison systems. These results demonstrate a consistent advantage on the evaluated in-house recognition sets, complementing the more varied rankings observed on public benchmarks.

\begin{table*}[!t]
\centering
\caption{Performance on in-house audio evaluations\textsuperscript{a}. ASR is evaluated using WER or CER (\%), S2TT using BLEU, and spoken interaction using the full-score rate (FSR, \%) defined in Eq.~\eqref{eq:dialogue_full_score_rate}. Bold indicates the best reported result for each evaluation; -- denotes an unreported result.}
\label{tab:spark-audio-inhouse-audio}
\footnotesize
\setlength{\tabcolsep}{2pt}
\renewcommand{\arraystretch}{1.08}
\begin{tabularx}{\linewidth}{@{}
>{\raggedright\arraybackslash}p{0.13\linewidth}
>{\raggedright\arraybackslash}p{0.23\linewidth}
>{\centering\arraybackslash}p{0.12\linewidth}
*{4}{>{\centering\arraybackslash}X}@{}}
\toprule
\textbf{Task} & \textbf{Benchmark / set} & \textbf{Metric} & \shortstack[c]{\textbf{Spark-Audio}\\\textbf{1.0}} & \shortstack[c]{\textbf{Qwen3.5-}\\ \textbf{Omni-Flash}} & \shortstack[c]{\textbf{Qwen3.5-}\\ \textbf{Omni-Plus}} & \shortstack[c]{\textbf{Gemini 3.1}\\\textbf{Pro}} \\
\midrule
Chinese ASR & General recognition & CER $\downarrow$ & \textbf{7.26} & 7.58 & 7.45 & 10.52 \\
\midrule
Dialect ASR & 202-Dialect corpus & CER $\downarrow$ & \textbf{14.29} & 20.53 & 19.79 & -- \\
\midrule
\multirow{2}{=}{\raggedright Multilingual ASR} & Core languages (14)\textsuperscript{b} & WER/CER $\downarrow$ & \textbf{7.75} & 11.41 & 10.86 & 13.37 \\
 & Extended languages (18)\textsuperscript{c} & WER/CER $\downarrow$ & \textbf{9.87} & 21.96 & 18.26 & 12.06 \\
\midrule
\multirow{2}{=}{\raggedright Spoken interaction\textsuperscript{d}} & General interaction & FSR $\uparrow$ & 63.28 & 49.57 & \textbf{65.36} & 58.76 \\
 & Instruction following & FSR $\uparrow$ & 67.92 & 60.83 & \textbf{70.42} & 69.58 \\
\midrule
\multirow{2}{=}{\raggedright S2TT} & zh$\rightarrow$en & BLEU $\uparrow$ & \textbf{44.76} & 32.05 & 36.12 & 37.35 \\
 & en$\rightarrow$zh & BLEU $\uparrow$ & \textbf{51.40} & 39.97 & 43.43 & 42.94 \\
\bottomrule
\end{tabularx}
\vspace{2pt}
\begin{minipage}{\linewidth}
\scriptsize
\setlength{\parindent}{0pt}
\textsuperscript{a}In-house evaluation sets are drawn from real-world application scenarios, such as meeting transcription and voice assistants.\par
\textsuperscript{b}The core language group is the same as that listed in Table~\ref{tab:spark-audio-public-audio}.\par
\textsuperscript{c}The extended group contains the 20 languages listed in Table~\ref{tab:spark-audio-public-audio}, excluding Tamil and Dutch because anomalous results were obtained for some comparison systems.\par
\textsuperscript{d}Responses are scored by a DeepSeek judge on a 0--5 scale. FSR is the percentage of responses receiving the maximum score of 5.
\end{minipage}
\end{table*}

\paragraph{Challenging Acoustic Conditions}
Table~\ref{tab:spark-audio-complex-asr} reports Chinese ASR results for high-noise and low-volume speech. Spark-Audio 1.0 obtains CERs of 37.76\% and 17.17\%, respectively, compared with 38.66\% and 23.62\% for Qwen3.5-Omni-Flash, the strongest reported comparison system in both conditions. The corresponding relative error reductions are 2.3\% and 27.3\%. The margin in the high-noise condition is small, whereas the margin on low-volume speech is substantial. The relatively high CERs in the high-noise condition indicate that substantial recognition errors remain on this test set.

\begin{table*}[!t]
\centering
\caption{In-house Chinese ASR performance under challenging acoustic conditions. CER is reported as a percentage; lower values are better. Bold indicates the best reported result for each condition; -- denotes an unreported result.}
\label{tab:spark-audio-complex-asr}
\footnotesize
\setlength{\tabcolsep}{2pt}
\renewcommand{\arraystretch}{1.08}
\begin{tabularx}{\linewidth}{@{}
>{\raggedright\arraybackslash}p{0.13\linewidth}
>{\raggedright\arraybackslash}p{0.23\linewidth}
>{\centering\arraybackslash}p{0.10\linewidth}
*{4}{>{\centering\arraybackslash}X}@{}}
\toprule
\textbf{Task} & \textbf{Benchmark / set} & \textbf{Metric} & \shortstack[c]{\textbf{Spark-Audio}\\\textbf{1.0}} & \shortstack[c]{\textbf{Qwen3.5-} \\ \textbf{Omni-Flash}} & \shortstack[c]{\textbf{Qwen3.5-}\\ \textbf{Omni-Plus}} & \shortstack[c]{\textbf{Gemini 3.1}\\\textbf{Pro}} \\
\midrule
\multirow{2}{=}{Chinese ASR} & High-noise speech & CER $\downarrow$ & \textbf{37.76} & 38.66 & 40.31 & -- \\
 & Low-volume speech & CER $\downarrow$ & \textbf{17.17} & 23.62 & 25.83 & 57.72 \\
\bottomrule
\end{tabularx}
\end{table*}

\subsection{Speech-to-Text Translation}
\label{subsec:s2tt_evaluation}

Translation results differ between the public and in-house evaluations. On the in-house sets, Spark-Audio 1.0 achieves 44.76 BLEU for Chinese-to-English translation and 51.40 BLEU for English-to-Chinese translation, exceeding the best comparison results by 7.41 and 7.97 BLEU points, respectively. On FLEURS, it obtains 28.40 and 44.54 BLEU in the same directions, below the corresponding best results of 32.20 and 47.80 BLEU from Qwen3.5-Omni-Plus. These results show that relative translation performance varies across evaluation corpora, with an advantage on the reported in-house sets and remaining gaps on FLEURS.

\subsection{Audio Understanding and Spoken Interaction}
\label{subsec:audio_understanding_evaluation}

On AIR-Bench, Spark-Audio 1.0 obtains a score of 71.85, close to Qwen3.5-Omni-Flash (72.63) but below Qwen3.5-Omni-Plus (79.23). On MMSU, it achieves 73.96\% accuracy, exceeding Qwen3.5-Omni-Flash (72.20\%) while trailing Qwen3.5-Omni-Plus (82.80\%) and Gemini 3.1 Pro (81.30\%). On MMAR, it obtains 57.90\% accuracy, below all three comparison systems. These results show that Spark-Audio 1.0 achieves comparable performance in audio perception and basic understanding (AIR-Bench and MMSU), whereas a gap remains in complex audio reasoning (MMAR).

For spoken interaction, Spark-Audio 1.0 achieves a score of 90.35 on VoiceBench, outperforming Qwen3.5-Omni-Flash (87.80) and Gemini 3.1 Pro (88.90), while remaining below Qwen3.5-Omni-Plus (93.10). On the in-house evaluations, it obtains full-score rates of 63.28\% for general interaction and 67.92\% for instruction following, exceeding Qwen3.5-Omni-Flash (49.57\% and 60.83\%) but remaining 2.08 and 2.50 percentage points below Qwen3.5-Omni-Plus, respectively. Gemini 3.1 Pro also achieves a higher instruction-following FSR (69.58\%). Overall, the interaction results support competitive performance in some settings while highlighting variability across benchmarks and interaction requirements.

\subsection{Text-Only Capabilities}
\label{subsec:text_capability_evaluation}

Table~\ref{tab:spark-audio-broader-capability} shows that Spark-Audio 1.0 performs comparably to Qwen3.5-Omni-Flash on the evaluated knowledge benchmarks: the absolute differences on MMLU-Pro, CMMLU, and C-Eval are all below 0.6 percentage points. Its scores are higher on GSM8K (97.35 versus 95.83) and HumanEval (92.07 versus 89.02), and lower on MATH-500 (96.60 versus 97.60). Instruction following remains a relative weakness, with scores of 82.19 on IFEval and 36.33 on IFBench, compared with 85.20 and 38.40 for Qwen3.5-Omni-Flash. These results establish competitive text-only performance on several evaluated tasks. 

\begin{table*}[!t]
\centering
\caption{Performance on public text-only benchmarks. All metrics are higher-is-better, and HumanEval is evaluated using pass@1. Bold indicates the better result for each benchmark.}
\label{tab:spark-audio-broader-capability}
\footnotesize
\setlength{\tabcolsep}{2pt}
\renewcommand{\arraystretch}{1.08}
\begin{tabularx}{\linewidth}{@{}
>{\raggedright\arraybackslash}p{0.20\linewidth}
>{\raggedright\arraybackslash}p{0.24\linewidth}
>{\centering\arraybackslash}p{0.12\linewidth}
*{2}{>{\centering\arraybackslash}X}@{}}
\toprule
\textbf{Task} & \textbf{Benchmark / set} & \textbf{Metric} & \shortstack[c]{\textbf{Spark-Audio}\\\textbf{1.0}} & \shortstack[c]{\textbf{Qwen3.5-}\\\textbf{Omni-Flash}} \\
\midrule
\multirow{3}{=}{\raggedright Knowledge} & MMLU-Pro & \mbox{Acc. $\uparrow$} & 79.52 & \textbf{79.90} \\
 & CMMLU & \mbox{Acc. $\uparrow$} & \textbf{82.39} & 81.85 \\
 & C-Eval & \mbox{Acc. $\uparrow$} & 85.85 & \textbf{86.00} \\
\midrule
\multirow{2}{=}{\raggedright Instruction following} & IFEval & \mbox{Acc.\textsuperscript{a} $\uparrow$} & 82.19 & \textbf{85.20} \\
 & IFBench & \mbox{Acc.\textsuperscript{a} $\uparrow$} & 36.33 & \textbf{38.40} \\
\midrule
\multirow{3}{=}{\raggedright Math \& Code} & MATH-500 & \mbox{Acc. $\uparrow$} & 96.60 & \textbf{97.60} \\
 & GSM8K & \mbox{Acc. $\uparrow$} & \textbf{97.35} & 95.83 \\
 & HumanEval & \mbox{pass@1 $\uparrow$} & \textbf{92.07} & 89.02 \\
\bottomrule
\end{tabularx}
\vspace{2pt}
\begin{minipage}{\linewidth}
\scriptsize
\setlength{\parindent}{0pt}
\textsuperscript{a}Using the strict prompt-level accuracy.
\end{minipage}
\end{table*}

%% file: sections/ablation.tex
\label{subsec:audio_encoder_ablation}

We evaluate the Spark-Audio encoder on speech recognition and audio understanding tasks. Specifically, we compare it with the Whisper large-v3 encoder~\citep{radford2023whisper} and the Qwen3-Omni Audio Transformer (AuT) encoder~\citep{xu2025qwen3}.

\paragraph{Experimental Setup}
Each encoder is paired with the same Qwen3-4B language model~\citep{yang2025qwen3}. Except for the starred multilingual AuT results, we freeze the encoder and language model, training only an adapter to map encoder representations into the language model's input space. All adapter-only experiments use the same training data (Chinese ASR, English ASR, multilingual ASR, and audio understanding), optimization settings, and trainable parameter budget. 

The available AuT encoder weights were jointly optimized with the Qwen3-Omni MoE backbone~\citep{xu2025qwen3}. When AuT is paired with Qwen3-4B, we find that adapter-only training does not fully reveal its multilingual capability. We therefore adopt a two-stage alignment for the starred FLEURS results: the first stage trains only the adapter, while the second stage jointly trains the encoder and adapter on the same data and fine-tunes the language model with LoRA.

\paragraph{Speech Recognition}
Table~\ref{tab:encoder-ablation-asr} shows the ASR results. In the adapter-only setting, the Spark-Audio encoder outperforms Whisper large-v3 on all 20 Chinese and English test sets and AuT on 19 of the 20, with improvements across both public benchmarks and in-house evaluation sets. It achieves average error rates of 8.93\% on the 15 Chinese test sets and 6.17\% on the five English test sets, compared with 10.47\% and 7.35\%, respectively, for AuT.

On FLEURS, the Spark-Audio encoder achieves a macro-averaged error rate of 26.24\% across 37 languages, an absolute reduction of 19.41 percentage points relative to Whisper large-v3 and 12.22 percentage points relative to AuT with two-stage adaptation. For the 16 languages that overlap with Qwen3-Omni's reported speech-input coverage, the average error rate is 16.40\%, corresponding to reductions of 6.99 and 2.99 percentage points, respectively. On the other 21 languages, it reduces the average error by 19.25 percentage points relative to AuT with two-stage adaptation. At the language level, the Spark-Audio encoder obtains lower error rates than Whisper on 36 of 37 languages and AuT on 34 of 37. These results demonstrate the advantage of the Spark-Audio encoder in multilingual speech recognition.

\begin{table*}[htbp]
\centering
\caption{Comparison of audio encoders on ASR benchmarks with Qwen3-4B as the language model backbone. We report WER or CER (\%). The lowest reported error in each row is shown in bold. Starred AuT results use two-stage adaptation; all other results use adapter-only training.}
\label{tab:encoder-ablation-asr}
\footnotesize
\setlength{\tabcolsep}{3pt}
\renewcommand{\arraystretch}{1.10}
\begin{tabularx}{\linewidth}{@{}
    >{\raggedright\arraybackslash}p{0.15\linewidth}
    >{\raggedright\arraybackslash}p{0.28\linewidth}
    >{\raggedright\arraybackslash}p{0.11\linewidth}
    *{3}{>{\centering\arraybackslash}X}@{}}
\toprule
\textbf{Task} &
\textbf{Dataset} &
\textbf{Metric} &
\shortstack[c]{\textbf{Spark-Audio}\\\textbf{Encoder}} &
\shortstack[c]{\textbf{Whisper}\\\textbf{large-v3}} &
\shortstack[c]{\textbf{Qwen3-Omni}\\\textbf{AuT encoder}} \\
\midrule
\multirow{4}{=}{\raggedright Chinese ASR}
    & AISHELL-1 & CER $\downarrow$ & \textbf{3.61} & 8.08 & 4.40 \\
    & AISHELL-2 & CER $\downarrow$ & \textbf{4.45} & 8.50 & 4.73 \\
    & In-house general (13 sets) & WER/CER $\downarrow$ & \textbf{9.68} & 36.06 & 11.38 \\
    & Average (15 sets) & WER/CER $\downarrow$ & \textbf{8.93} & 32.35 & 10.47 \\
\midrule
\multirow{4}{=}{\raggedright English ASR}
    & LibriSpeech test-clean & WER $\downarrow$ & \textbf{3.63} & 5.15 & 4.76 \\
    & LibriSpeech test-other & WER $\downarrow$ & \textbf{6.77} & 10.84 & 9.24 \\
    & In-house domains (3 sets) & WER $\downarrow$ & \textbf{6.82} & 10.51 & 7.59 \\
    & Average (5 sets) & WER $\downarrow$ & \textbf{6.17} & 9.50 & 7.35 \\
\midrule
\multirow{3}{=}{\raggedright Multilingual ASR}
    & FLEURS (overlap, 16 languages) & WER/CER $\downarrow$ & \textbf{16.40} & 23.39 & 19.39\textsuperscript{*} \\
    & FLEURS (other, 21 languages) & WER/CER $\downarrow$ & \textbf{33.74} & 62.61 & 52.99\textsuperscript{*} \\
    & FLEURS (all 37 languages) & WER/CER $\downarrow$ & \textbf{26.24} & 45.65 & 38.46\textsuperscript{*} \\
\bottomrule
\end{tabularx}
\vspace{2pt}
\begin{minipage}{\linewidth}
\scriptsize
\setlength{\parindent}{0pt}
FLEURS results are macro-averaged over languages. The 16-language overlap group comprises languages included in Qwen3-Omni's reported speech-input coverage: Arabic, German, Spanish, French, Indonesian, Italian, Japanese, Korean, Malay, Dutch, Portuguese, Russian, Thai, Turkish, Urdu, and Vietnamese. The other 21 languages are Bulgarian, Bengali, Czech, Greek, Persian, Filipino, Hausa, Hindi, Javanese, Kazakh, Khmer, Lao, Mongolian, Burmese, Polish, Romanian, Swedish, Swahili, Tamil, Ukrainian, and Uzbek.
\textsuperscript{*}\,The AuT encoder multilingual results use the two-stage setup described above. All other results use adapter-only training.
\end{minipage}
\end{table*}

\paragraph{Audio Understanding}
Table~\ref{tab:encoder-ablation-understanding} shows the audio understanding results. The Spark-Audio encoder achieves an average accuracy of 61.49\% across eight evaluation sets, improving over Whisper large-v3 and AuT by 28.00 and 6.95 percentage points, respectively. It outperforms Whisper on all eight sets and achieves the highest or joint-highest accuracy on six of the eight.

The advantage over AuT is particularly pronounced on the four benchmarks covering sounds and acoustic scenes: ESC-50~\citep{piczak2015esc50}, NonSpeech7K~\citep{rashid2023nonspeech7k}, CochlScene~\citep{jeong2022cochlscene}, and TUT~\citep{mesaros2016tut}. Across these benchmarks, the Spark-Audio encoder outperforms AuT by 16.72 percentage points in average accuracy. On age prediction, it outperforms AuT by 13.50 percentage points. Performance is less uniform on emotion and gender prediction: the Spark-Audio encoder matches AuT on the in-house emotion set at 30.77\%, but trails it by 7.50 percentage points on CASIA and 17.30 percentage points on gender prediction. This pattern shows that the gains vary across tasks, with the clearest advantages over AuT in sound understanding, scene understanding, and age prediction.

\begin{table*}[htbp]
\centering
\caption{Comparison of audio encoders on audio understanding tasks. We report accuracy (\%). The highest accuracy in each row is shown in bold, including ties.}
\label{tab:encoder-ablation-understanding}
\footnotesize
\setlength{\tabcolsep}{3pt}
\renewcommand{\arraystretch}{1.10}
\begin{tabularx}{\linewidth}{@{}
    >{\raggedright\arraybackslash}p{0.25\linewidth}
    >{\raggedright\arraybackslash}p{0.18\linewidth}
    >{\raggedright\arraybackslash}p{0.11\linewidth}
    *{3}{>{\centering\arraybackslash}X}@{}}
\toprule
\textbf{Task} &
\textbf{Dataset} &
\textbf{Metric} &
\shortstack[c]{\textbf{Spark-Audio}\\\textbf{Encoder}} &
\shortstack[c]{\textbf{Whisper}\\\textbf{large-v3}} &
\shortstack[c]{\textbf{Qwen3-Omni}\\\textbf{AuT encoder}} \\
\midrule
\multirow{2}{=}{\raggedright Emotion}
    & CASIA & Acc. $\uparrow$ & 25.00 & 11.67 & \textbf{32.50} \\
    & In-house set & Acc. $\uparrow$ & \textbf{30.77} & 25.38 & \textbf{30.77} \\
\midrule
Environmental sound
    & ESC-50 & Acc. $\uparrow$ & \textbf{59.50} & 25.25 & 31.50 \\
Non-speech sound
    & NonSpeech7K & Acc. $\uparrow$ & \textbf{76.83} & 47.72 & 63.31 \\
\multirow{2}{=}{\raggedright Acoustic scene}
    & CochlScene & Acc. $\uparrow$ & \textbf{70.66} & 42.70 & 64.54 \\
    & TUT & Acc. $\uparrow$ & \textbf{61.54} & 16.67 & 42.31 \\
Age
    & In-house set & Acc. $\uparrow$ & \textbf{89.10} & 38.00 & 75.60 \\
Gender
    & In-house set & Acc. $\uparrow$ & 78.50 & 60.55 & \textbf{95.80} \\
\midrule
Average & All 8 sets & Acc. $\uparrow$ & \textbf{61.49} & 33.49 & 54.54 \\
\bottomrule
\end{tabularx}
\end{table*}

Overall, these results demonstrate that the Spark-Audio encoder is an effective front end for a general audio language model. Its strong multilingual speech recognition performance and gains on sound, scene, and age prediction show that the learned representations can effectively capture both semantic and acoustic information.

%% file: sections/conclusion.tex
In this paper, we present Spark-Audio 1.0, a speech foundation model trained entirely on domestic AI accelerators. It supports a broad range of tasks, including multilingual speech recognition, multi-dialect speech recognition, speech translation, audio perception and understanding, and text- and speech-based dialogue. Notably, it supports speech recognition in 99 languages and 202 dialects. The model couples a 0.65B-parameter dense audio encoder with a 30B-A3B Mixture-of-Experts language model. The encoder is pretrained with representation-enhanced BEST-RQ and then refined through multi-task supervised training. We evaluate Spark-Audio 1.0 on public benchmarks as well as in-house test sets derived from real-world application scenarios. Trained on 13 million hours of audio data with extensive text data, Spark-Audio 1.0 is competitive with the similarly sized Qwen3.5-Omni-Flash on Chinese–English speech-to-text translation, audio perception, and text-based dialogue (knowledge, math, and code), and outperforms it on multilingual speech recognition and spoken interaction. It shows particularly clear advantages on realistic, application-oriented tasks. Spark-Audio 1.0 also leaves room for improvement, particularly in instruction following and complex audio reasoning. In future work, we will explore more effective use of audio data and broaden coverage across acoustic conditions and application domains. We will also extend the model to hour-long speech recognition, multi-speaker speech recognition, and spatially aware audio understanding.

%% file: sections/authors.tex
\textbf{Contributors}\footnotemark[1]:

\begin{tabular}{lllll}
Chenyue Zhang & Cong Liu        & Dingshu Tian        & Dong Sun             & Genshun Wan \\
Guanming Meng  & Guolong Zhong  & Hang Chen          & Hankun Yang\textsuperscript{\#}   & Haoming Wang  \\
Haoyang Wang   &  Hengquan Mei   & Huaxin Wu         & Jiahao Li     & Jiahuan Fan\textsuperscript{\#}  \\
Jiamin Ma      & Jianqing Gao      &  Jigang Tang\textsuperscript{\#}  & Jin Liu       & Jingchen Li   \\
Jinjie Fu\textsuperscript{\#}      & Kaixuan Guo          & Kun Song       &  Li Xu     & Li Zhang      \\
Lijuan Liu    & Liyan Luo           & Ming Sheng           & Mingen Zhao    &  Mingting Hu  \\
Ningqian Wu   & Pan Xu        & Pei Xu              & Qiangyu Tan          & Rui Wang\textsuperscript{\#}  \\
Ruzi Li      & Saier Jin     & Shibo Xu      & Shifu Xiong$^{*}$   & Shihao Jiang  \\
Tan Liu        & Wei Wu       & Wei Xu        & Wenyu Wang    & Xiangyu Jia         \\
Xiao Li       & Xiaowen Duan   & Xiaoxue Guo  & Xin Chen      & Xinyi Liang   \\
Xu Chen       & Yanan Song    &   Yang Xu        & Yexia Huang  & Yibo Li\textsuperscript{\#}  \\
Yifan Xu\textsuperscript{\#}  & Yikang Liu         & Yikun Li   &   Yiting Lin     & Yi Fang    \\
Yuxin Wang    & Zan Feng       & Zhenchuan Li       & Zhuo Chen\textsuperscript{\#}  & Zichao Zhang   \\
Ziwen Wang\textsuperscript{\#}   
\end{tabular}

\vspace{0.5em}
Names marked with \textsuperscript{\#} denote individuals who have departed from our team. We sincerely appreciate their contributions to this work.
\footnotetext[1]{Alphabetical order.}
\begingroup
\renewcommand{\thefootnote}{}
\footnotetext[1]{$^*$ Corresponding author (email: sfxiong@iflytek.com).}
\endgroup

\clearpage